\documentclass[proof]{pasj00}
\draft

\SetRunningHead{Aoki et al.}{GRB~060206}
\Received{2008/12/01}
\Accepted{}
\Published{}

\begin{document}
\title{No Evidence for Variability of Intervening Absorption Lines toward
GRB 060206: Implications for the Mg II Incidence Problem
\thanks{Based on data collected at Subaru Telescope, which is operated by the
National Astronomical Observatory of Japan.}}
\author{
Kentaro \textsc{Aoki},\altaffilmark{1}
Tomonori \textsc{Totani},\altaffilmark{2}
Takashi \textsc{Hattori},\altaffilmark{1}
Kouji \textsc{Ohta},\altaffilmark{2}\\
Koji S. \textsc{Kawabata},\altaffilmark{3}
Naoto \textsc{Kobayashi},\altaffilmark{4}
Masanori \textsc{Iye},\altaffilmark{5}
Ken'ichi \textsc{Nomoto},\altaffilmark{6}
\and
Nobuyuki \textsc{Kawai}\altaffilmark{7}\\
}

\altaffiltext{1}{Subaru Telescope, National Astronomical Observatory of Japan,
\\  650 North A`ohoku Place, Hilo, HI 96720, U.S.A.}
\altaffiltext{2}{Department of Astronomy, Kyoto University,
\\Sakyo-ku, Kyoto 606-8502}
\altaffiltext{3}{Hiroshima Astrophysical Science Center, Hiroshima University,
\\  1-31-1 Kagamiyama, Higashi-Hiroshima, Hiroshima 739-8526}
\altaffiltext{4}{Institute of Astronomy, University of Tokyo,
\\  2-21-1, Osawa, Mitaka, Tokyo}
\altaffiltext{5}{National Astronomical Observatory of Japan,
\\  2-21-1, Osawa, Mitaka, Tokyo}
\altaffiltext{6}{Institute for the Physics and Mathematics of the Universe,
\\University of Tokyo, 5-1-5 Kashiwanoha, Kashiwa, Chiba 277-8568}
\altaffiltext{7}{Department of Physics, Tokyo Institute of Technology,
\\  2-12-1 Ookayama, Meguro-ku, Tokyo 152-8551}

\maketitle

\begin{abstract}
We examine variability of absorption line strength of intervening
systems along the line of sight to GRB 060206 at $z = 4.05$, by the
low-resolution optical spectra obtained by the Subaru telescope from
six to ten hours after the burst. 
Strong variabilities of
Fe~\emissiontype{II} and Mg~\emissiontype{II} lines at $z=1.48$
during $t=$5--8 hours have been reported for this GRB \citep{Hao07},
and this has been used to support the idea of clumpy Mg~\emissiontype{II} 
cloudlets that was originally proposed to explain the anomalously high
incidence of Mg~\emissiontype{II} absorbers in GRB spectra compared with quasars.
However, our spectra with higher signal-to-noise ratio do not show any
evidence for variability in $t=$6-10 hours.  
There is a clear
discrepancy between our data and Hao et al. data in the overlapping
time interval.  Furthermore, the line strengths in our data are in
good agreement with those observed at $t \sim$ 2 hours by
\citet{Tho08}.  We also detected Fe~\emissiontype{II} and
Mg~\emissiontype{II} absorption lines for a system at $z = 2.26$,
and these lines do not show evidence for variability either.  
Therefore we conclude  
that there is no strong evidence for variability
of intervening absorption lines toward GRB 060206,
significantly weakening the support to the Mg~\emissiontype{II} cloudlet hypothesis 
by the GRB~060206 data.
\end{abstract}

\section{Introduction}

The gamma-ray burst (GRB) has opened up a new window to probe high
redshift universe, now reaching comparable redshifts with the studies
by galaxies and quasars, and starting to probe the era of the cosmic
reionization \citep{Kawai06}.  The optical GRB afterglow spectra give
us a lot of information about GRB host galaxies \citep{Mir02, Sav03,
  Prochaska07, Chen07}, intervening absorption systems along the line
of sight \citep{Proch06, Sud07, Tej07} and the reionization of the
universe \citep{Tot06}.

GRB~060206 is the 10th highest redshift GRB as of 2008 November.  GRB~060206
was triggered by the $Swift$ Burst Alert Telescope on 2006 February 6 at
04:46:53 UT.  The redshift of the burst was determined to be 4.048
\citep{Fyn06}.  The intervening absorption lines are reported at
$z=1.48$ \citep{Fyn06,Hao07,Tho08} and $z=$2.26 (Aoki et al. 2006;
this work).  Surprisingly, \citet{Hao07} claimed strong variation of
the equivalent widths (EWs) of Fe~\emissiontype{II} $\lambda 2600.2$
and Mg~\emissiontype{II} $\lambda2803.5$ at $z$=1.48; the EW of the
Fe~\emissiontype{II} $\lambda 2600.2$ system at $z=1.48$ showed a more
than 80\% decrease within 1.5 hours duration. It is difficult to
interpret this system to be located in the GRB host galaxy, because a
large redshift difference requires relativistic motion and hence a
large velocity dispersion of lines is expected, while the absorption
line widths are only $14 - 20$ km s$^{-1}$ \citep{Tho08}.  Then,
almost unique possible interpretation is that the intervening
absorption system is smaller than the image size of the GRB afterglow,
and the line strength changes with the size evolution of the GRB
afterglow. This requires the size of the absorption system to be less
than $\sim 10^{16}$ cm \citep{Hao07}.

This variability, if real, would change the standard picture of
intervening absorption systems found in bright cosmological objects
such as quasars and GRBs. Furthermore, this result has been used to
support the Mg~\emissiontype{II} cloudlet hypothesis that has been
proposed to explain the anomalous Mg~\emissiontype{II} incidence along
the GRB sight lines \citep{Frank07}. \citet{Proch06}
reported that the incidence of Mg~\emissiontype{II} absorption lines
along GRB sight lines is about four times higher than that for
quasars, at a statistical significance greater than 99.9\% confidence limit, and
this cannot easily be explained by known selection effects.  If the
intervening line variability of GRB 060206 is in fact due to the small
size of the Mg~\emissiontype{II} cloudlets, the anomalous incidence of
Mg~\emissiontype{II} systems may also be explained by the larger image
size of quasars than GRB afterglows.

However, GRB 060206 is currently the only one case for the intervening
line variability, and it should be examined carefully.  Here we report
the analysis of the optical afterglow spectra of GRB 060206 taken by
the Subaru telescope during $t = $ 6.01--9.73 hours after the burst, a
part of which is overlapping with the Hao et al.'s observation ($t = $
4.13--7.63 hours).  We will examine especially the line variability of
the intervening Mg~\emissiontype{II} and Fe~\emissiontype{II} systems.
In addition to the $z=1.48$ system, we will give detailed analysis
of the $z=2.26$ system as well and examine the variability. 

 When we completed most of this paper, the revised version of
  Th\"one et al. (2008, arXiv:0708.3448v2) appeared on the preprint
  server, in which they also analyzed the same data from the Subaru
  archive and reached a similar conclusion of no evidence for
  variability. However, details of the Subaru observation and data
  analysis are not given in their paper. Here we give more detailed descriptions
  about the observation logs and the processing of the Subaru data
  taken by ourselves, as well as the implications for the
  Mg~\emissiontype{II} incidence problem.  The $z=2.26$ system was
  originally reported by Aoki et al. (2006) and it is mentioned also
  in Fynbo et al. (2006), but Th\"one et al. (2008) concluded that
  this is not real. However, we will show that this system is real,
  based on the detection of seven Fe~\emissiontype{II} lines and Mg II
  doublet.

\section{Observations and Data Reduction} 
\label{Obs} 

The afterglow of GRB~060206 was observed on 2006 February 6 (UT) with
FOCAS \citep{Kashik02} attached to the Subaru 8.2-m telescope
\citep{Iye04}.  We obtained eight spectra of 30 minutes integration
for each at 10:47, 11:19, 11:51, 12:23, 12:54, 13:28, 13:59, and 14:31
(UT).  All these times are midpoint of exposures.  We used the R300
grism with two different order-cut filters for the observations.  The
first three and last two spectra were taken with the O58 filter which
covers between 5700 \AA~and 10200 \AA.  The other three spectra were
taken with the Y47 filter which allows an uncontaminated spectrum
between 4800 \AA~and 9200 \AA.  The slit width was set to be
\timeform{0.8''}.  The atmospheric dispersion corrector was used and
the slit position angle was \timeform{176D}.  The spectrophotometric
standard star Hz~44 was observed for sensitivity calibration and
removal of atmospheric absorption lines.  It was photometric condition
with good seeing (\timeform{0.6''} - \timeform{0.9''} in $R-$band).
The seeing size became smaller as the GRB was rising, and it was
smaller than the slit width at the final epoch of the observations.
Consequently, the resolution of the GRB's spectra was changed from
8.0\AA~to 6.8~{\AA} (290 km s$^{-1}$ to 250 km s$^{-1}$ at 8190 \AA),
which is confirmed by measuring absorption lines of a
bright star found in the same slit for the GRB afterglow.
\par
The data were reduced using IRAF\footnote {IRAF is distributed by the
  National Optical Astronomy Observatory, which is operated by the
  Association of Universities for Research in Astronomy (AURA), Inc.,
  under cooperative agreement with the National Science Foundation.  }
for the procedures of bias subtraction, flat-fielding, wavelength
calibration, and sky subtraction.  Wavelength calibration was
performed using OH night sky emission lines, and the rms wavelength
calibration error is 0.2 - 0.3 \AA.  The sensitivity calibration was
performed as a function of wavelength, and the atmospheric absorption
feature was removed by using the spectrum of Hz~44.  Fringe pattern
exists beyond 7700 \AA~ in the spectra, and fringe spectra were made
from the standard star's spectra normalized by its continuum.  We
divided the afterglow spectra by the fringe spectra, and suppressed
the fringe pattern.  The initial strength of fringe pattern is 6--11\%
of the continuum level of the afterglow spectra, but it is reduced by
a factor of $\sim$2 by this process.  The foreground Galactic
extinction of $A_{B}=0.054$ mag \citep{Schle98} was corrected.

\section{Results} \label{Results}

\begin{figure}
  \begin{center}
    \FigureFile(135mm,100mm){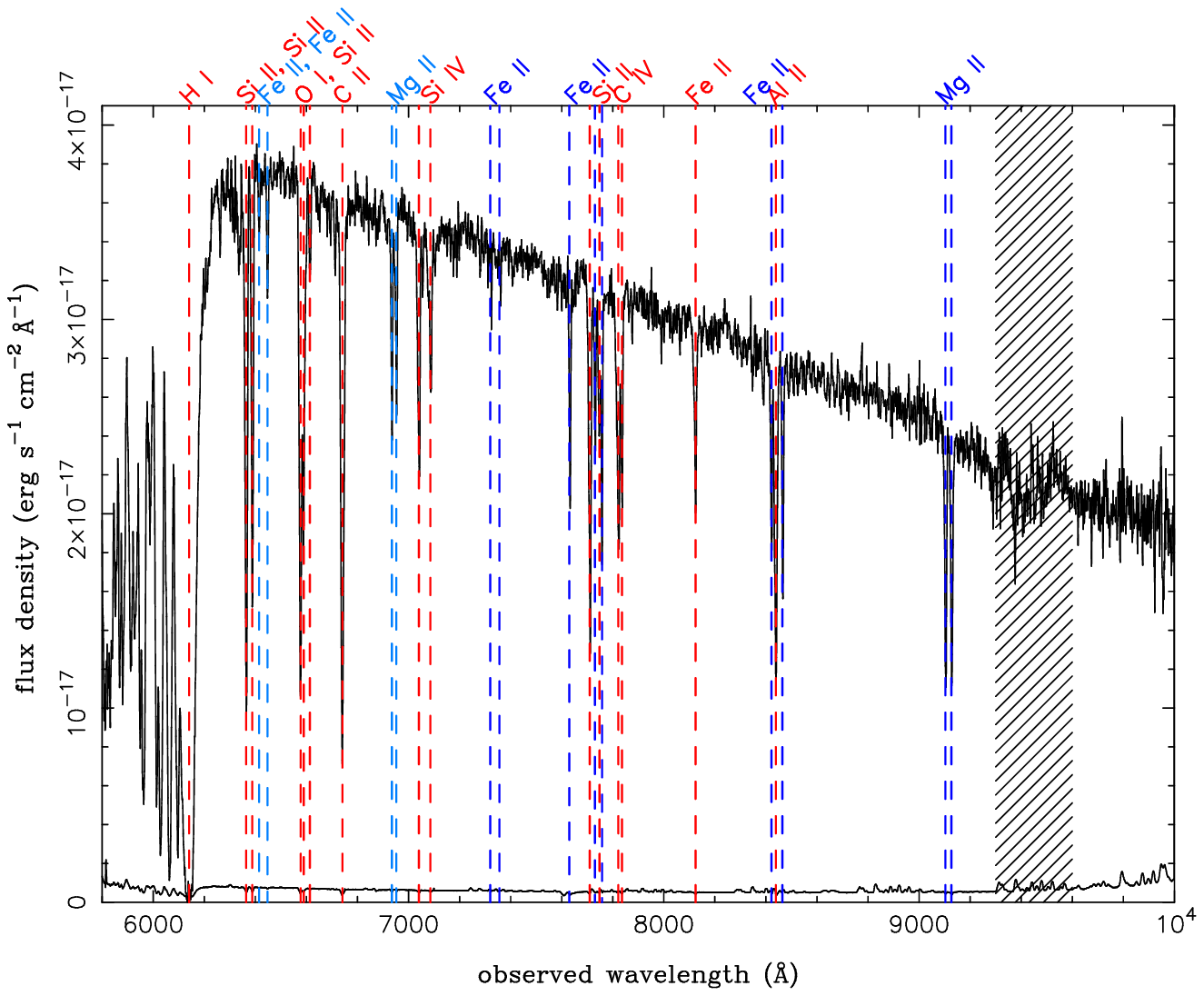}
  \end{center}
  \caption{
The observed spectrum of GRB~060206 afterglow in vacuum wavelength.
This spectrum was taken at 9.7 hours after the burst.
Red, cyan, and blue dashed lines indicate absorption lines at $z=4.05, 1.48$ and $2.26$,
respectively.
The hatched region between 9300 and 9600 \AA~indicates the place of water vapor absorption lines in
the atmosphere.
The spectrum at the bottom is the noise spectrum.
}
\label{fig1}
\end{figure}

Figure \ref{fig1} displays the optical spectrum of GRB~060206 taken
9.7 hours after the burst.  The absorption feature between 9300 and
9600 \AA~is residuals of the telluric absorption removal.  The
spectrum clearly shows the $z=4.05$ and $1.48$ absorption line systems
which have been reported by \citet{Fyn06}, \citet{Hao07}, and
\citet{Tho08}.  Furthermore, the $z=2.26$ absorption line system of Mg
~\emissiontype{II} is clearly recognized.  We also detected seven
Fe~\emissiontype{II} resonance lines ($\lambda2249.9, 2260.8, 2344.3, 2374.5,
2382.8, 2586.7$, and $2600.2$) at $z=2.26$.  
All these $z=2.26$
absorption lines are beyond 7200 {\AA} and hence outside of the
coverage of the WHT/ISIS observation \citep{Fyn06,Tho08}, explaining
that \citet{Tho08} could not find the absorption system at $z=2.26$.
We list all detected lines in table \ref{tblA} in Appendix.
In figure \ref{fig11} in Appendix, we show the spectrum combined all 4 hours data. 
\par
\begin{figure}
  \begin{center}
    \FigureFile(135mm,110mm){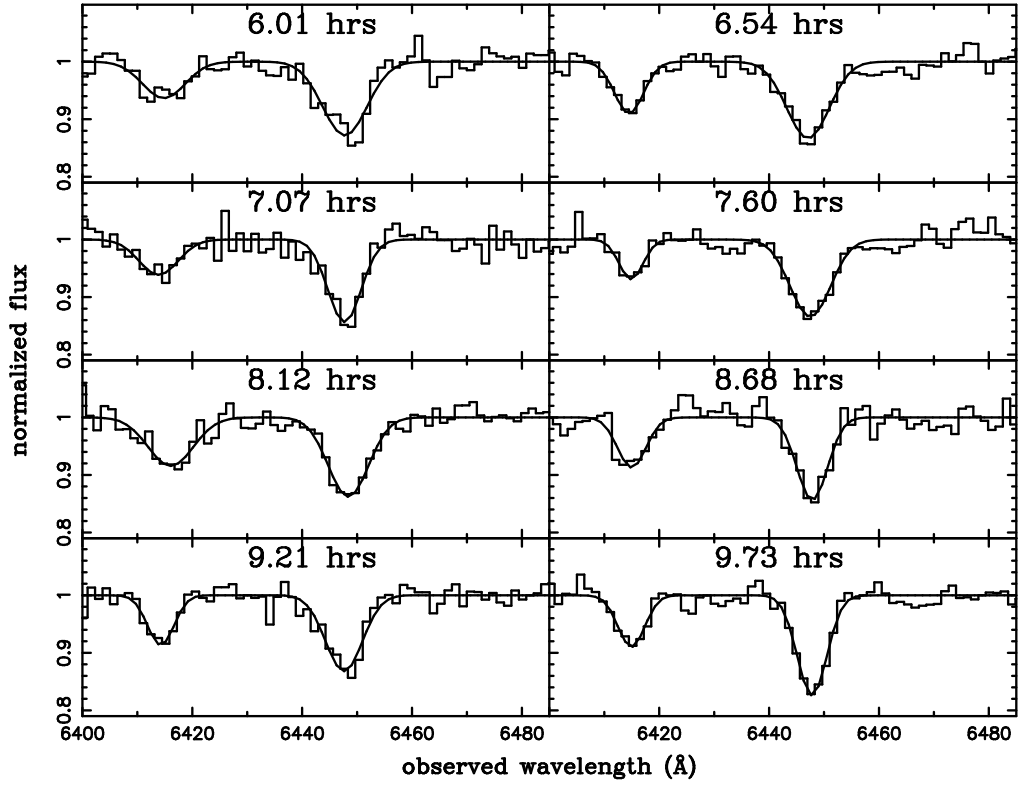}
  \end{center}
  \caption{Spectra of GRB~060206 afterglow around
    Fe~\emissiontype{II} $\lambda\lambda2586.7, 2600.2$ at $z=1.48$.
    All spectra are normalized by the continuum, and all absorption
    lines are fitted with a Gaussian.}
\label{fig2}
\end{figure}
\begin{figure}
  \begin{center}
    \FigureFile(135mm,110mm){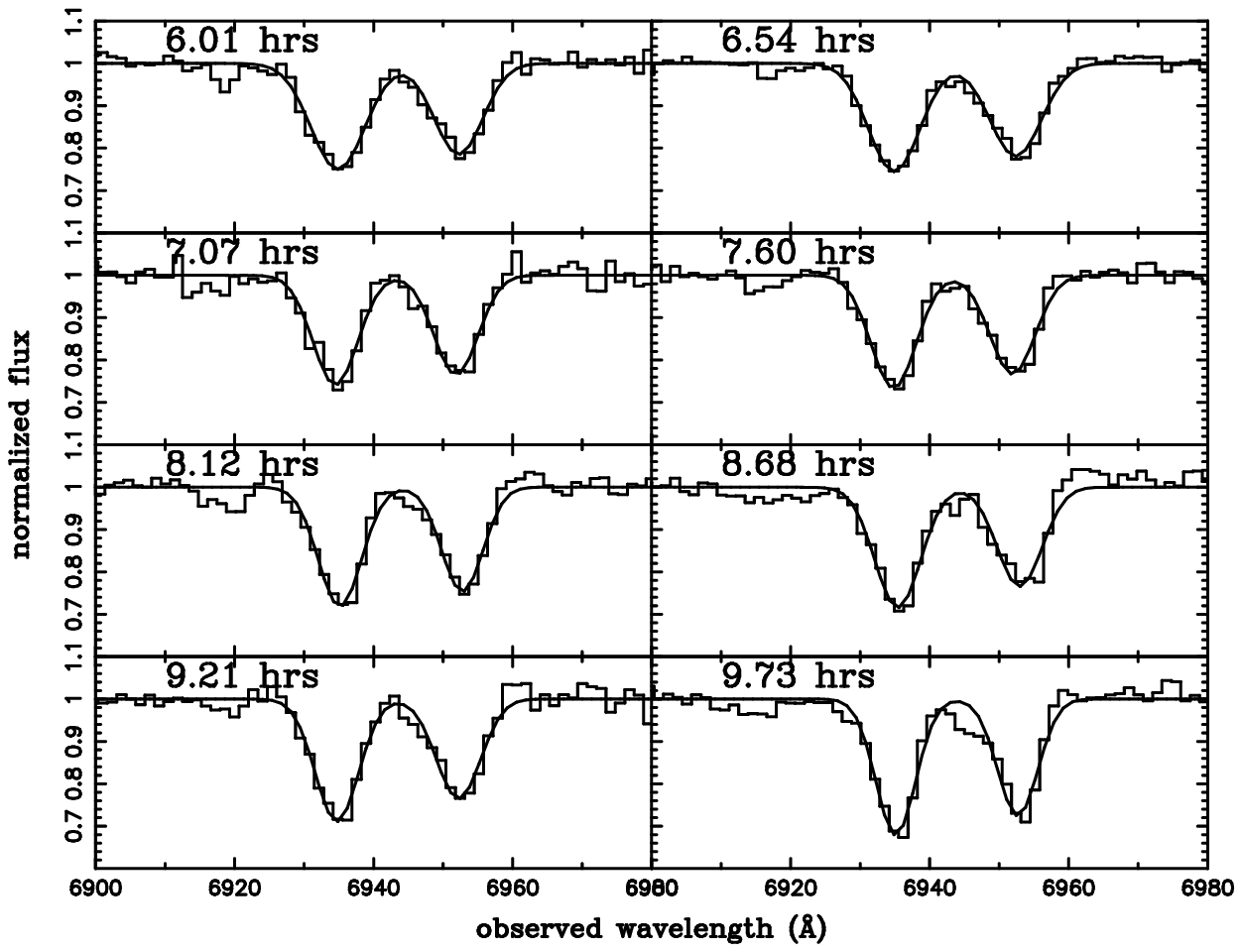}
  \end{center}
  \caption{The same as Fig. 2, but for
    Mg~\emissiontype{II} $\lambda\lambda2796.4, 2803.5$ at $z=1.48$.  }
\label{fig3}
\end{figure}

We show Fe~\emissiontype{II} $\lambda \lambda 2586.7, 2600.2$
absorption line profiles at $z=1.48$ in our normalized spectra in
Figure \ref{fig2}.  No significant temporal variation is seen between
6.0 and 9.7 hours after the burst.  The absorption lines at 9.73 hours
after the burst are narrower than those at other epochs, but this is
due to the change of the spectral resolution; note that seeing size is
smaller than the slit width at this epoch, as mentioned in Section 2.
We also show Mg~\emissiontype{II} $\lambda\lambda2796.4,2803.5$
absorption lines at $z=1.48$ in Figure \ref{fig3}.  No significant
temporal variation is seen, either.  We measured EW in the observer's
frame for these lines by fitting each absorption line with a single
Gaussian.  The Gaussian fits are shown in Figure 2 and 3, and the EW
results are tabulated in table \ref{tbl1}.  Figure 4 presents the time
evolution of EW of these lines, and we confirm that there is no
significant variation in the strength of all the four $z=1.48$
absorption lines.
The difference between the first and last observations is 0.14 \AA~at 
maximum (Mg~\emissiontype{II} $\lambda\lambda2796.4$), and that is
smaller than $1\sigma$ uncertainty.
The EW measurements of Mg~\emissiontype{II}
$\lambda 2803.5$, Fe~\emissiontype{II} $\lambda$2600.2, and
Fe~\emissiontype{II} $\lambda$2586.7 in the same Subaru data performed
by Th\"one et al. (2008) are mostly consistent with our
measurements.  However, we find that the EW of Mg~\emissiontype{II}
$\lambda 2796.4$ in Th\"one et al. (2008) is significantly smaller than our
measurement.  The reason for this discrepancy is not clear, although
the conclusions of no evidence for variability are the same.

The absorption line profiles of Fe~\emissiontype{II}
$\lambda\lambda2586.7, 2600.2$, Mg~\emissiontype{II}
$\lambda\lambda2796.4,2803.5$, 
Fe~\emissiontype{II} $\lambda\lambda2249.9, 2260.8$,
Fe~\emissiontype{II} $\lambda2344.2$,
and Fe~\emissiontype{II} $\lambda\lambda2374.5, 2382.8$,
for another intervening absorption line
system at $z=2.26$ are shown in Figure \ref{fig5}, \ref{fig6},
\ref{fig7}, \ref{fig8} and \ref{fig9},
respectively. The absorption line at 8439 \AA~ is
Al~\emissiontype{II} at $z=4.05$, and 
those at 7710 \AA~and 7748 \AA~are Si~\emissiontype{II} $\lambda1526.7$ and
Si~\emissiontype{II}* $\lambda1533.4$ at $z=4.05$, respectively. 
The equivalent widths are measured
in the same way as the system at $z=1.48$, and the results are listed
in table \ref{tbl2}.  We do not find any systematic change of
equivalent width in these all absorption lines at $z=2.26$ but one, either
(Figure \ref{fig10}).  
The EW of Fe~\emissiontype{II} $\lambda2382.8$ changed more than $3\sigma$
between 6.0542 hours and 7.065 hours.
Since the oscillator strength of Fe~\emissiontype{II} $\lambda2374.5$ ($3.13\times10^{-2}$) 
is smaller than that of Fe~\emissiontype{II} $\lambda2382.8$ ($3.20\times10^{-1}$),
Fe~\emissiontype{II} $\lambda2374.5$ is less saturated, and thus
should be more variable than Fe~\emissiontype{II} $\lambda2382.8$.
However, the EW of Fe~\emissiontype{II} $\lambda2374.5$ was stable within $1\sigma$.
As seen in Figure \ref{fig9}, Fe~\emissiontype{II} $\lambda2382.8$ was heavily blended with Si~\emissiontype{II}* $\lambda1533.4$ at $z=4.05$.
This may cause larger uncertainty in EW, and large variation in 
Fe~\emissiontype{II} $\lambda2382.8$ is doubtful. 
Note that Mg~\emissiontype{II} $\lambda2803.5$
is systematically stronger than Mg~\emissiontype{II} $\lambda2796.4$
in Figure \ref{fig7}, although the strength of Mg~\emissiontype{II}
$\lambda2803.5$ should be the same (optically thick case) as, or
weaker (optically thin case) than, Mg~\emissiontype{II}
$\lambda2796.4$.  
We noticed Si~\emissiontype{II} $\lambda1808.01$ at $z=4.05$ is probably overlapped with Mg~\emissiontype{II} $\lambda2803.5$ at 9130.5 \AA.
Using the column density of Si\emissiontype{II} ($1.7\times10^{15}$) from \citet{Tho08},
we estimated EW of Si~\emissiontype{II} $\lambda1808.01$ to be
0.56 \AA~in observers' frame (optically thin case).
We confirm that Si~\emissiontype{II} $\lambda1808.01$ absorption line will be optically thin with the column density, the Doppler parameter $b$ of 20 ${\rm km s}^{-1}$ and the small oscillator strength ($2.08\times 10^{-3}$).
We consider that the contamination of Si~\emissiontype{II} $\lambda1808.01$ is likely
the cause of a systematic increase of Mg~\emissiontype{II} $\lambda2803.5$ in this result.

\section{Discussion and Conclusion}

\begin{figure}
  \begin{center}
    \FigureFile(130mm,100mm){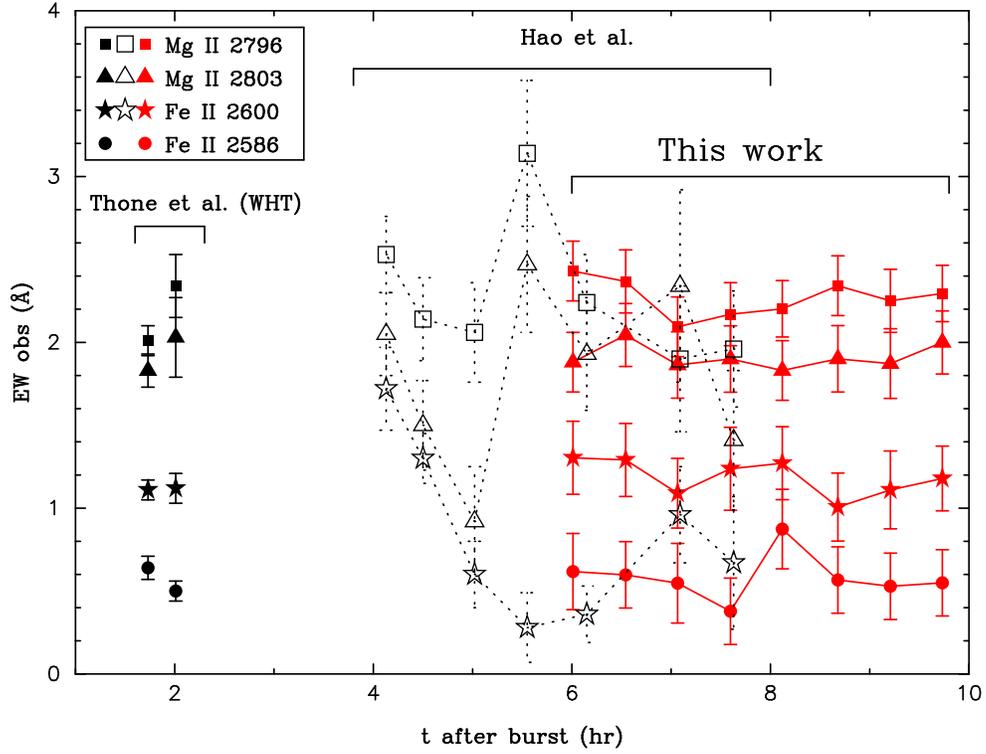}
  \end{center}
  \caption{Equivalent width (observer's frame) of the $z=1.48$
    absorption lines versus observation time after the burst.  The EWs
    of Fe~\emissiontype{II} $\lambda\lambda2586.7, 2600.2$, and
    Mg~\emissiontype{II} $\lambda\lambda2796.4, 2803.5$ are shown with
    the symbols indicated in the figure.  Our measurements are
    indicated by filled red symbols between 6 hours and 10 hours, and
    Hao et al. (2007) data by open ones between 4 hours and 8 hours.
    The filled black symbols at 2 hours are from Th\"one et
    al. (2008).  }
\label{fig4}
\end{figure}

\begin{figure}
  \begin{center}
    \FigureFile(135mm,110mm){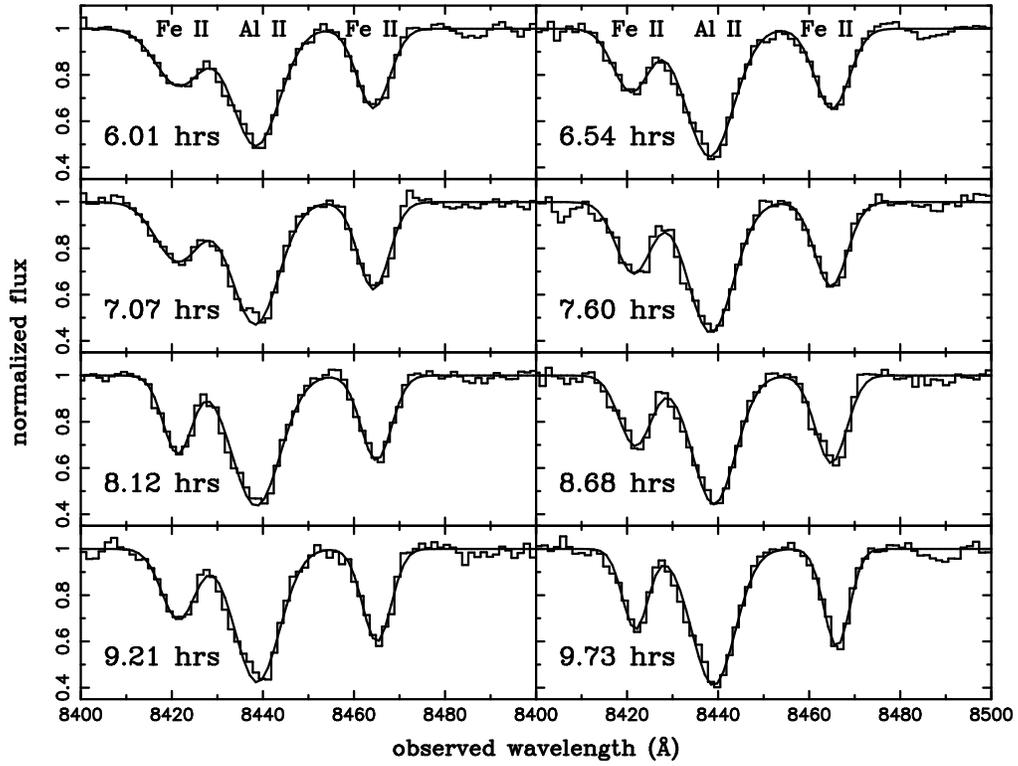}
  \end{center}
  \caption{The same as Fig. 2, but for
    Fe~\emissiontype{II} $\lambda\lambda2586.7, 2600.2$ at $z=2.26$.
    The absorption line at 8439 \AA~is Al~\emissiontype{II} at
    $z=4.05$.  }
\label{fig5}
\end{figure}

\begin{figure}
  \begin{center}
    \FigureFile(135mm,110mm){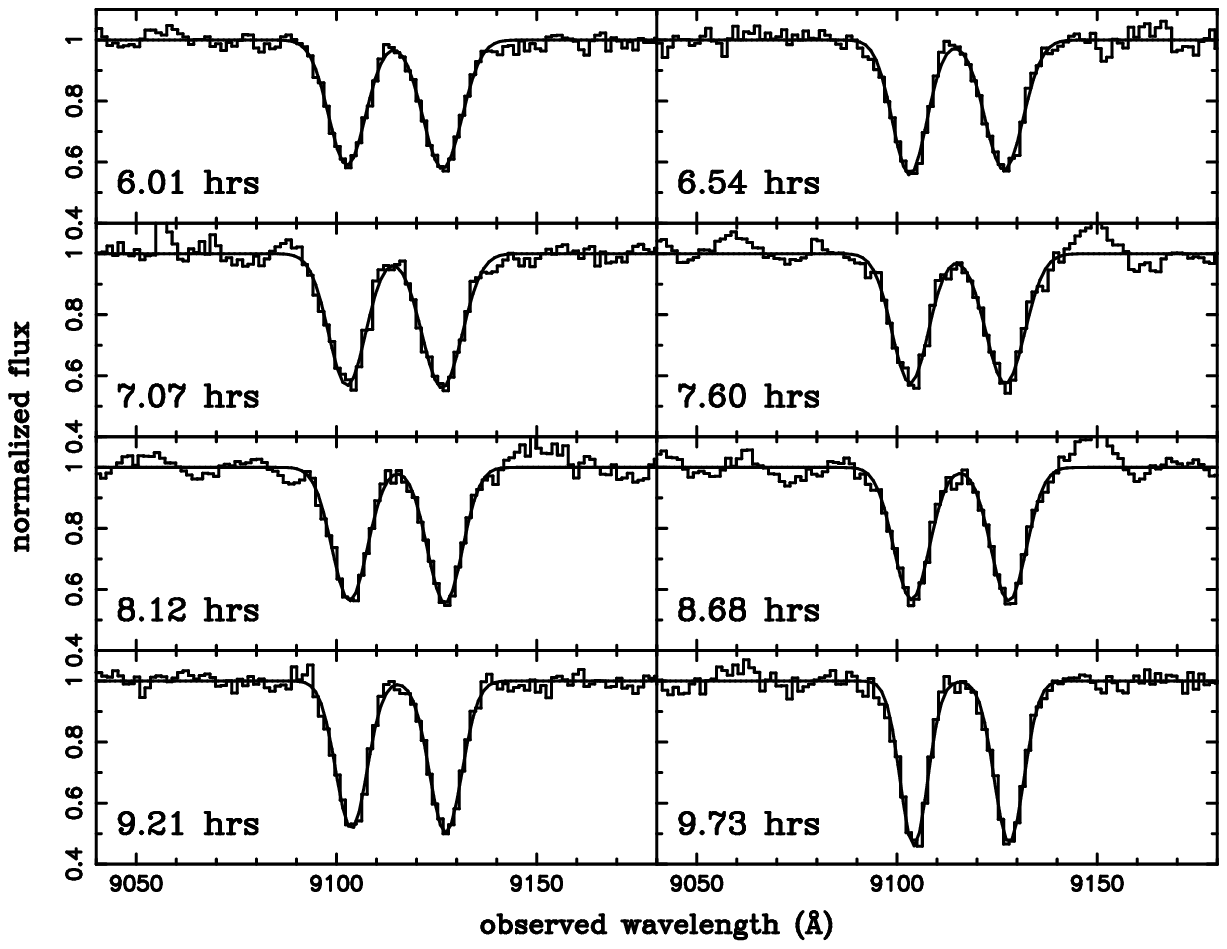}
  \end{center}
  \caption{The same as Fig. 2, but for Mg~\emissiontype{II} $\lambda\lambda2796.4,2803.5$ at $z=2.26$.
}
\label{fig6}
\end{figure}

\begin{figure}
  \begin{center}
    \FigureFile(135mm,110mm){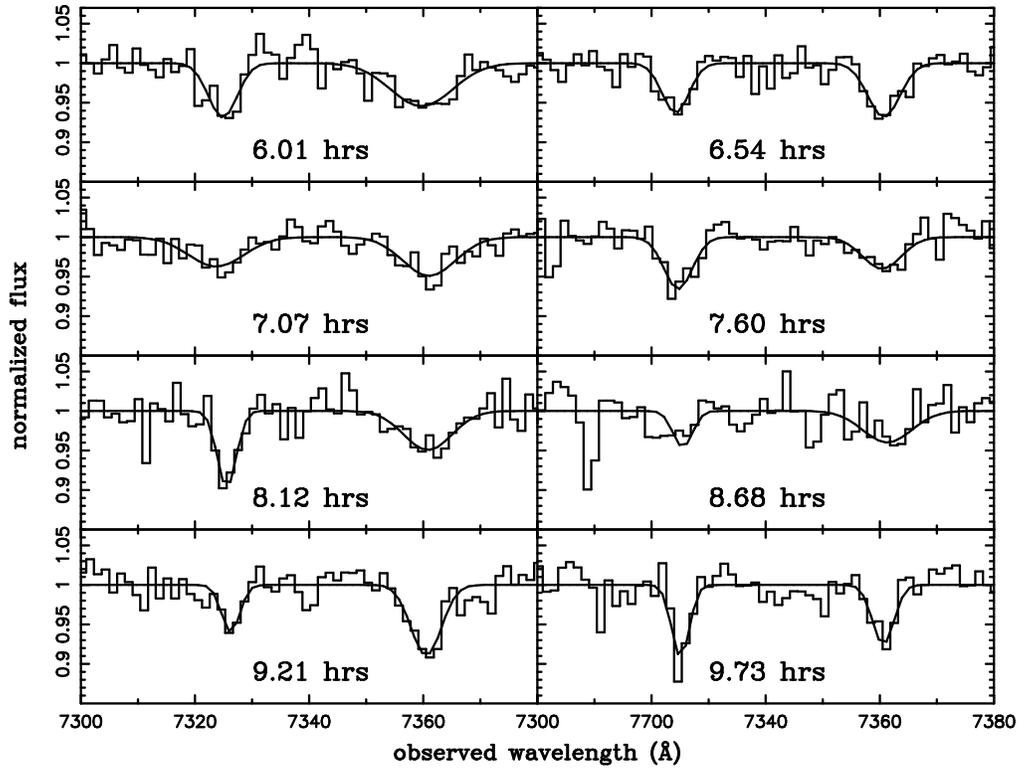}
  \end{center}
  \caption{The same as Fig. 2, but for Fe~\emissiontype{II} $\lambda\lambda2249.9, 2260.8$ at $z=2.26$.
}
\label{fig7}
\end{figure}

\begin{figure}
  \begin{center}
    \FigureFile(135mm,110mm){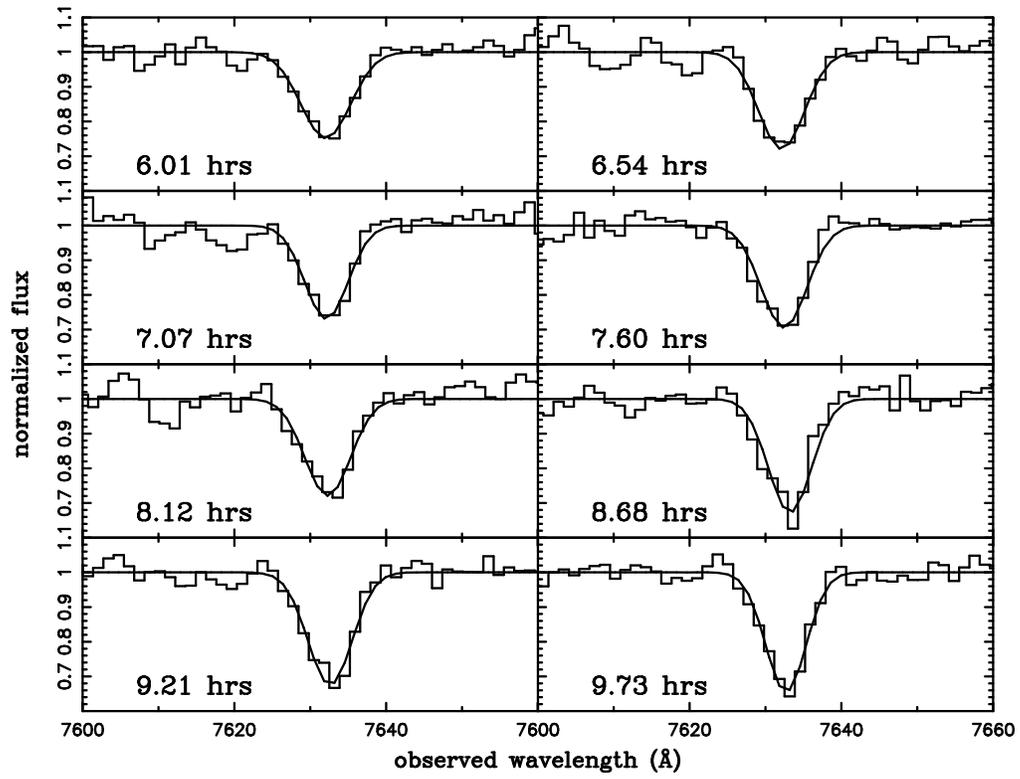}
  \end{center}
  \caption{The same as Fig. 2, but for Fe~\emissiontype{II} $\lambda2344.2$ at $z=2.26$.
}
\label{fig8}
\end{figure}

\begin{figure}
  \begin{center}
    \FigureFile(135mm,110mm){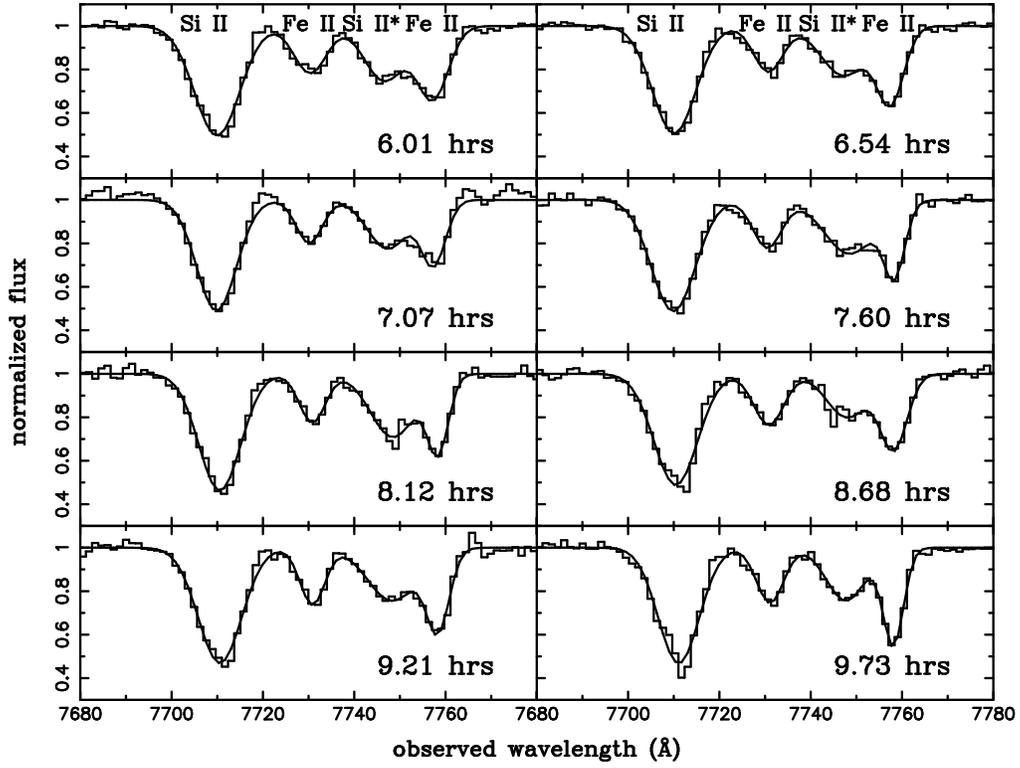}
  \end{center}
  \caption{The same as Fig. 2, but for Fe~\emissiontype{II} $\lambda\lambda2374.5, 2382.8$ at $z=2.26$.
The absorption lines at 7710 \AA~and 7748 \AA~are Si~\emissiontype{II} $\lambda1526.7$ and
Si~\emissiontype{II}* $\lambda1533.4$ at $z=4.05$, respectively. 
}
\label{fig9}
\end{figure}

\begin{figure}
  \begin{center}
    \FigureFile(130mm,100mm){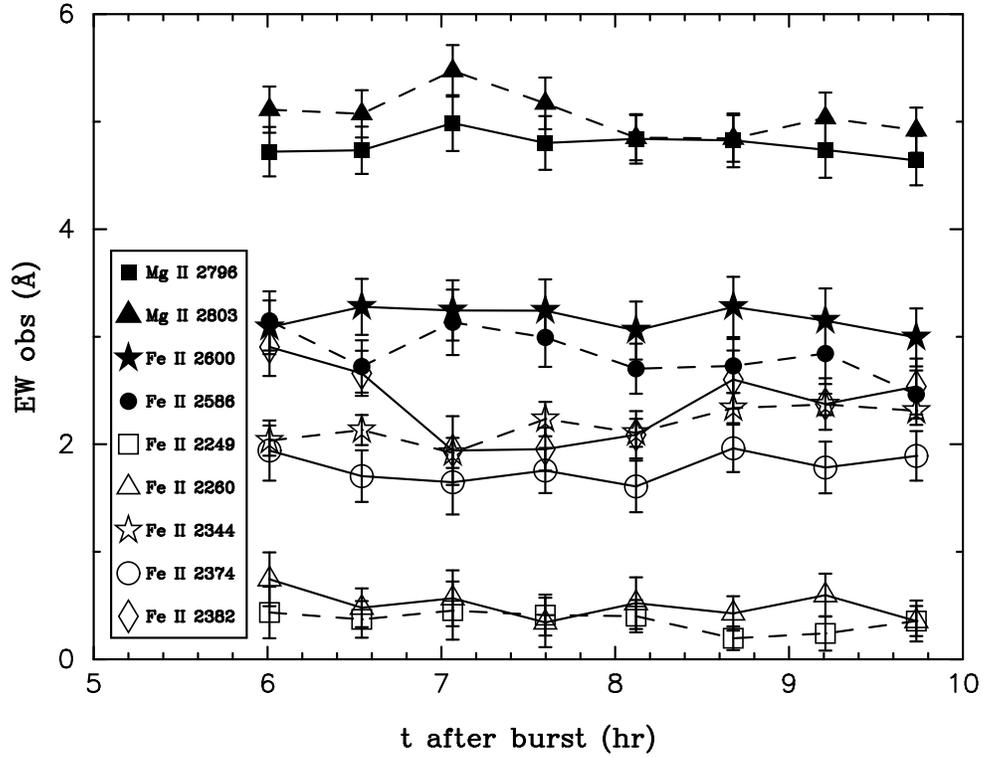}
  \end{center}
  \caption{Equivalent widths (observer's frame) of the $z=2.26$ 
Mg~\emissiontype{II} $\lambda2796.4$ (filled square), 
Mg~\emissiontype{II} $\lambda2803.5$ (filled triangle),
Fe~\emissiontype{II} $\lambda2600.2$ (filled star), 
Fe~\emissiontype{II} $\lambda2586.7$ (filled circle),
Fe~\emissiontype{II} $\lambda2249.9$ (open square), 
Fe~\emissiontype{II} $\lambda2260.8$ (open triangle),
Fe~\emissiontype{II} $\lambda2344.2$ (open star), 
Fe~\emissiontype{II} $\lambda2374.5$ (open circle), and
Fe~\emissiontype{II} $\lambda2382.8$ (open diamond)
absorption lines, as functions of time after the burst.}
\label{fig10}
\end{figure}

In addition to the Subaru data, the equivalent width measurements of
the $z = 1.48$ system by \citet{Hao07} and the WHT/ISIS data of
\citet{Tho08} are shown in Figure \ref{fig4}.  Our early four spectra
are overlapping in time with the last three observations by
\citet{Hao07}.  Our EW measurement of $z=1.48$ Fe~\emissiontype{II}
$\lambda2600.2$ at $t = $ 6.011 hours is clearly in disagreement with
the value measured by \citet{Hao07} at $t= $ 6.15 hours.  It should be
noted that Fe~\emissiontype{II} $\lambda2586.7$ at $z=1.48$ is clearly
detected at 6415 \AA~ in our spectra (Figure \ref{fig2}) with an
equivalent width of $\sim 0.55$ \AA.  On the other hand, it is not
mentioned by \citet{Hao07}, and in fact, this line cannot be clearly
seen in the Hao et al.'s spectra.  This indicates that an EW of
$\lesssim 0.5$ \AA~ cannot reliably be measured by the Hao et al.'s
observation, and hence the claim of the variability of
Fe~\emissiontype{II} $\lambda2600.2$ line should suffer from a large
uncertainty.

The Hao et al.'s spectra show a strong variability of
Mg~\emissiontype{II} lines around $t = 5 - 5.5$ hours, and we cannot
examine this variability because this epoch is not covered by our
data. However, the early measurements of EWs of these lines at $z =
1.48$ by \citet{Tho08} at $t \sim 2$ hours are in good agreement with
our measurements. It seems rather unlikely that the lines show a
variability only in the time duration of the Hao et al.'s observation
and no variability at all in the earlier or later data with higher
signal-to-noise ratio.  Furthermore, there is no evidence for time
evolution of EWs in another absorption system at $z = 2.26$.  From
these results, we conclude that the variability reported by
\citet{Hao07} is likely due to some artificial effects or
statistical fluke in low signal-to-noise data.

Our result implies that the data of GRB 060206 no longer provide a
strong support to the Mg~\emissiontype{II} cloudlet hypothesis proposed by
\citet{Frank07} for the anomalously high incidence of
Mg~\emissiontype{II} systems in GRB spectra.  Furthermore, several
difficulties of this hypothesis have been pointed out.  One such
argument is that apparently unsaturated Mg~\emissiontype{II}
absorption lines with the doublet ratio of 1:1 are never found in
intervening absorption lines, although such lines are expected if
optically thick Mg~\emissiontype{II} cloudlets partially cover quasar
beams \citep{Proch06, Porc07, Sud07}.  Another argument is that there
is no apparent change in Mg~\emissiontype{II} incidence over the
continuum and broad emission lines of quasar spectra, although the
broad line regions of quasars are much larger than the continuum
emitting region \citep{Pont07}.  It should also be noted that the
suggested size of the Mg~\emissiontype{II} cloudlets is several orders
of magnitude smaller than those derived from the observations of
gravitationally lensed quasars \citep{Rau02, Ell04}.
\par
Therefore, we consider that the Mg~\emissiontype{II} incidence anomaly
is likely to be explained by some other effects.  Although the anomaly has
not yet been solved, it could be explained, or at least the
statistical significance of the anomaly is reduced, by a combination
of some selection effects such as dust extinction and gravitational
lensing \citep{Porc07, Sud07}.

\bigskip We are grateful to the Subaru Telescope staff for their
assistance during our observations.  This work was partly supported by
the Grant-in-Aid for Scientific Research on Priority Areas (19047003)
from the Ministry of Education, Culture, Sports, Science and
Technology (MEXT) of Japan.  

\appendix
\section*{Absorption Lines List}
\begin{figure}
  \begin{center}
    \FigureFile(135mm,100mm){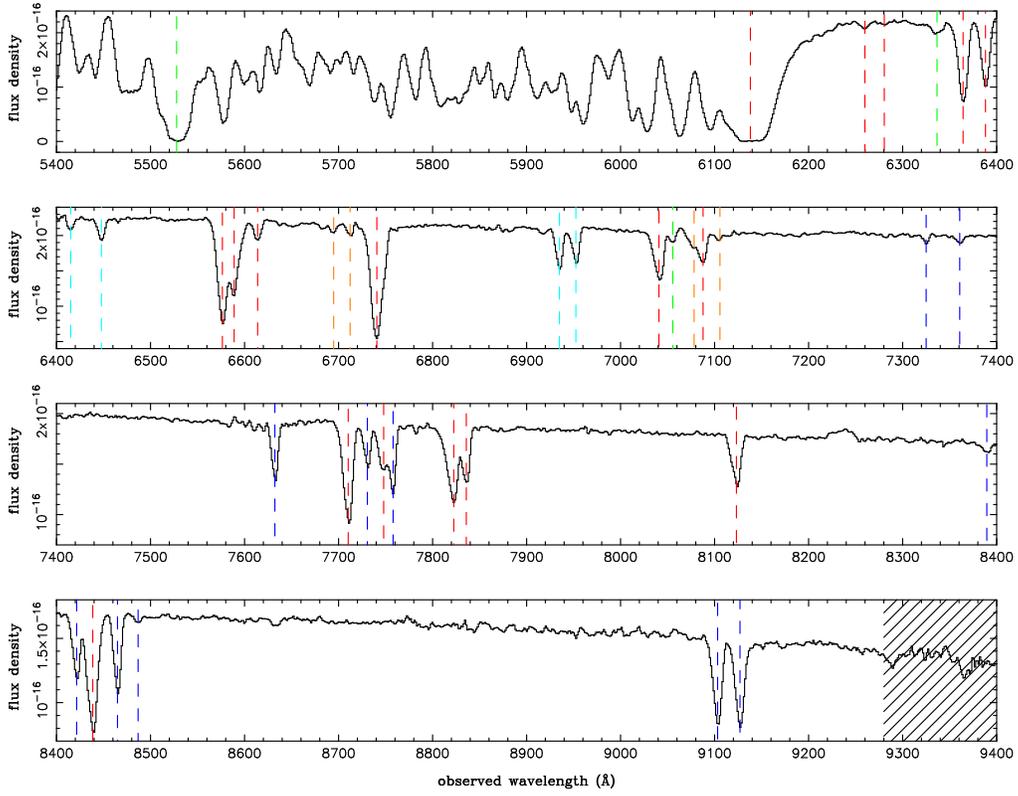}
  \end{center}
  \caption{
The observed spectrum of GRB~060206 afterglow in vacuum wavelength.
This high signal-to-noise ratio spectrum was made by combining all data.
The hatched region between 9280 and 9600 \AA~ indicates the place of water vapor absorption lines in
the atmosphere.
Red, green, blue and cyan dashed lines indicate absorption lines at 
$z=4.05, 3.55, 2.26$, and 1.48, respectively.
The unidentified lines are indicated by orange dashed lines.
See identifications in Table \ref{tblA}.
}
\label{fig11}
\end{figure}

We identified absorption lines in the spectrum combined all 4 hrs data (Figure \ref{fig11}),
but 1.5 hrs integration between 4800 \AA~and 6000 \AA,
and 2.5 hrs between 8800 \AA~and 10200 \AA.
The identifications, observed wavelengths in vacuum, and equivalent widths in
rest frame are listed in
Table~\ref{tblA}.
Most absorption lines redward of 7000 \AA~in GRB~060206 are reported here at the first time

\clearpage

\clearpage

\clearpage

\clearpage

\clearpage

\clearpage

\clearpage

\clearpage

\clearpage

\clearpage

\clearpage

\clearpage

\begin{table}
  \caption{Equivalent width (observer's frame)
of the $z=1.48$  Absorption Lines.
}\label{tbl1}
  \begin{center}
\begin{tabular}{ccccc}
\hline
Hour & EW$_{\rm Fe II 2600}$ & EW$_{\rm Fe II 2586}$ & EW$_{\rm Mg II 2796}$ &EW$_{\rm Mg II 2803}$ \\
\hline
6.011 & $1.30\pm0.22 $ & $0.62\pm0.23 $ & $2.43\pm0.18 $ & $1.88\pm 0.18 $\\
6.542 & $1.29\pm0.22 $ & $0.60\pm0.20 $ & $2.37\pm0.19 $ & $2.04\pm 0.19 $\\
7.065 & $1.09\pm0.21 $ & $0.55\pm0.24 $ & $2.09\pm0.18 $ & $1.86\pm 0.20 $\\
7.598 & $1.24\pm0.25 $ & $0.38\pm0.20 $ & $2.17\pm0.19 $ & $1.90\pm 0.20 $\\
8.120 & $1.27\pm0.22 $ & $0.87\pm0.24 $ & $2.20\pm0.17 $ & $1.83\pm 0.18 $\\
8.679 & $1.01\pm0.21 $ & $0.57\pm0.20 $ & $2.34\pm0.18 $ & $1.90\pm 0.20 $\\
9.209 & $1.11\pm0.24 $ & $0.53\pm0.20 $ & $2.25\pm0.19 $ & $1.87\pm 0.21 $\\
9.732 & $1.18\pm0.20 $ & $0.55\pm0.20 $ & $2.29\pm0.17 $ & $2.00\pm 0.19 $\\
\hline
\end{tabular}
\end{center}
\end{table}
\clearpage

\begin{table}
  \caption{Equivalent width (observer's frame)
of the $z=2.26$  Absorption Lines.}\label{tbl2}
  \begin{center}
\begin{tabular}{cccccc}
\hline
Hour & EW$_{\rm Fe II 2600}$ & EW$_{\rm Fe II 2586}$ & EW$_{\rm Mg II 2796}$ &EW$_{\rm Mg II 2803}$ \\
\hline
6.011 & $3.09\pm0.25$ & $3.15\pm0.28$ & $4.72\pm0.23$  & $5.11\pm0.22$ & \\
6.542 & $3.28\pm0.26$ & $2.72\pm0.24$ & $4.73\pm0.22$  & $5.07\pm0.22$ & \\
7.065 & $3.24\pm0.28$ & $3.13\pm0.30$ & $4.99\pm0.26$  & $5.47\pm0.24$ & \\
7.598 & $3.24\pm0.29$ & $2.99\pm0.27$ & $4.80\pm0.25$  & $5.17\pm0.24$ & \\
8.120 & $3.06\pm0.27$ & $2.70\pm0.23$ & $4.84\pm0.23$  & $4.85\pm0.21$ & \\
8.679 & $3.28\pm0.28$ & $2.73\pm0.25$ & $4.83\pm0.25$  & $4.84\pm0.22$ & \\ 
9.209 & $3.15\pm0.30$ & $2.84\pm0.28$ & $4.74\pm0.26$  & $5.04\pm0.24$ & \\ 
9.732 & $2.99\pm0.27$ & $2.46\pm0.22$ & $4.64\pm0.23$  & $4.92\pm0.21$ & \\ 
\hline
\hline
Hour & EW$_{\rm Fe II 2249}$ & EW$_{\rm Fe II 2260}$ & EW$_{\rm Fe II 2344}$ &EW$_{\rm Fe II 2374}$ & EW$_{\rm Fe II 2382}$ \\
\hline
6.011 & $0.44\pm0.24$ & $0.74\pm0.25$ & $2.0\pm0.14$  & $1.9\pm0.28$ & $2.9\pm0.27$ \\
6.542 & $0.37\pm0.17$ & $0.48\pm0.18$ & $2.1\pm0.14$  & $1.7\pm0.24$ & $2.7\pm0.21$ \\
7.065 & $0.45\pm0.27$ & $0.57\pm0.26$ & $1.9\pm0.14$  & $1.6\pm0.30$ & $1.9\pm0.32$ \\
7.598 & $0.41\pm0.19$ & $0.34\pm0.23$ & $2.2\pm0.16$  & $1.8\pm0.21$ & $ 2.0\pm0.20$ \\
8.120 & $0.40\pm0.15$ & $0.52\pm0.24$ & $2.1\pm0.13$  & $1.6\pm0.24$ & $2.1\pm0.22$ \\
8.679 & $0.20\pm0.11$ & $0.43\pm0.16$ & $2.3\pm0.14$  & $2.0\pm0.22$ & $2.6\pm0.27$ \\
9.209 & $0.24\pm0.16$ & $0.60\pm0.20$ & $2.4\pm0.10$  & $1.8\pm0.24$ & $2.4\pm0.24$ \\
9.732 & $0.36\pm0.19$ & $0.36\pm0.14$ & $2.3\pm0.13$  & $1.9\pm0.23$ & $2.5\pm0.26$ \\
\hline
\end{tabular}
\end{center}
\end{table}
\clearpage

\begin{longtable}{ccccc}
  \caption{Line Identification}\label{tblA}
  \hline
  Wavelength (\AA) & Line & $z$ & $W_{r}$ (\AA)& Ref. \\
  \hline
\endhead
\hline
\multicolumn{5}{@{}l@{}}{\hbox to 0pt{\parbox{180mm}{\footnotesize
1: Hao et al. 2007, 2: Fynbo et al. 2006, 3: Th\"one et al. 2008
}}}
\endfoot
\hline
\endlastfoot
5527.8 & H \emissiontype{I} 1215.67       & 3.5471 & --    & 1    \\
6138.1 & H \emissiontype{I} 1215.67       & 4.0492 & --    & 2 \\
6259.9 & N \emissiontype{V} 1238.82       & 4.0531 & $0.093\pm0.015$ & 3  \\
6280.4 & N \emissiontype{V} 1242.80       & 4.0534 & $0.063\pm0.017$ & 3  \\
6336.5 & Si \emissiontype{IV} 1393.76     & 3.5463 & $0.46\pm0.048$  & 1    \\
6364.3 & Si \emissiontype{II} 1260.42     & 4.0494 & $1.6\pm0.017$   & 3  \\
6388.1 & Si \emissiontype{II}* 1264.74, 1265.00 & 4.0504 & $1.1\pm0.016$   & 3  \\
6415.1 & Fe \emissiontype{II} 2586.65     & 1.4801 & $0.20\pm0.027$  & 3  \\
6447.8 & Fe \emissiontype{II} 2600.17     & 1.4798 & $0.48\pm0.033$  & 1, 3    \\
6576.5 & O \emissiontype{I} 1302.17       & 4.0504 & $1.5\pm0.023$ & 3  \\
6589.0 & O \emissiontype{I}* 1304.86, 1306.03, Si \emissiontype{II} 1304.37     & 4.0487 & $1.1\pm0.028$   & 3  \\
6614.0 & Si \emissiontype{II}* 1309.28    & 4.0516 & $0.20\pm0.012$ & 3  \\
6694.8 & ? & -- & --  &        \\
6712.3 & ? & -- & -- &        \\
6740.8 & C \emissiontype{II} 1334.53, C \emissiontype{II}* 1335.66, 1335.71    & 4.0482 & $2.2\pm0.019$ & 3  \\
6935.0 & Mg \emissiontype{II} 2796.35     & 1.4800 & $0.93\pm0.031$ & 1,  3  \\
6952.4 & Mg \emissiontype{II} 2803.53     & 1.4799 & $0.77\pm0.031$ & 1, 3   \\
7040.8 & Si \emissiontype{IV} 1393.76     & 4.0517 & $0.79\pm0.017$  &      \\
7055.6 & C \emissiontype{IV} 1550.77      & 3.5497 & $0.092\pm0.011$ &      \\
7078.2 & ? & -- & --  &      \\
7087.7 & Si \emissiontype{IV} 1402.77     & 4.0527 & $0.34\pm0.051$ &      \\
7105.7 & ? & -- & -- &  \\
7325.0 & Fe \emissiontype{II} 2249.88     & 2.2557 & $0.10\pm0.0089$ &      \\
7360.7 & Fe \emissiontype{II} 2260.78     & 2.2558 & $0.14\pm0.012$  &      \\
7632.3 & Fe \emissiontype{II} 2344.21     & 2.2558 & $0.68\pm0.023$  &      \\
7710.3 & Si \emissiontype{II} 1526.71     & 4.0503 & $1.2\pm0.024$  &      \\
7730.6 & Fe \emissiontype{II} 2374.46     & 2.2557 & $0.55\pm0.029$ &      \\
7748.0 & Si \emissiontype{II}* 1533.43    & 4.0527 & $0.63\pm0.047$ &      \\
7758.0 & Fe \emissiontype{II} 2382.77     & 2.2559 & $0.74\pm0.061$ &      \\
7823.3 & C \emissiontype{IV} 1548.20      & 4.0532 & $0.69\pm0.15$  &      \\
7835.8 & C \emissiontype{IV} 1550.77      & 4.0528 & $0.51\pm0.016$  &      \\
8123.2 & Fe \emissiontype{II} 1608.45     & 4.0503 & $0.59\pm0.012$ &      \\
8389.5 & Mn \emissiontype{II} 2576.88     & 2.2557 & $0.17\pm0.024$  &      \\
8421.4 & Fe \emissiontype{II} 2586.65     & 2.2557 & $0.89\pm0.023$  &      \\
8438.6 & Al \emissiontype{II} 1670.79     & 4.0507 & $1.4\pm0.017$ &      \\
8464.9 & Fe \emissiontype{II} 2600.17     & 2.2555 & $0.98\pm0.020$  &      \\
8486.8 & Mn \emissiontype{II} 2606.46     & 2.2561 & $0.083\pm0.018$ &      \\
9103.2 & Mg \emissiontype{II} 2796.35     & 2.2554 & $1.5\pm0.029$ &      \\
9127.0 & Mg \emissiontype{II} 2803.53     & 2.2555 & $1.6\pm0.029$ &      \\
\end{longtable}
\clearpage

\end{document}